\documentclass[a4paper,twocolumn,11pt,unpublished]{quantumarticle}
\pdfoutput=1

\usepackage[bookmarks=true,breaklinks]{hyperref}
\usepackage{graphicx,color}
\usepackage[english]{babel}
\usepackage{mathtools}
\usepackage{physics}
\usepackage{tikz}
\usetikzlibrary{quantikz}

\begin{document}

\title{On quantum factoring using noisy intermediate scale quantum computers}

\author{Vivian Phan}
\email{vivian.phan@aalto.fi}
\affiliation{Micro and Quantum Systems group\\
Department of Electronics and Nanoengineering\\
Aalto University}

\author{Arttu Pönni}
\email{arttu.ponni@aalto.fi}
\affiliation{Micro and Quantum Systems group\\
Department of Electronics and Nanoengineering\\
Aalto University}

\author{Matti Raasakka}
\email{matti.raasakka@aalto.fi}
\affiliation{Micro and Quantum Systems group\\
Department of Electronics and Nanoengineering\\
Aalto University}

\author{Ilkka Tittonen}
\email{ilkka.tittonen@aalto.fi}
\affiliation{Micro and Quantum Systems group\\
Department of Electronics and Nanoengineering\\
Aalto University}

\begin{abstract}
We study the performance and resource usage of the variational quantum factoring (VQF) algorithm for different instance sizes and optimization algorithms. Our simulations show better chance of finding the ground state when using VQE rather than QAOA for optimization. In gradient-based optimization we find that the time required for quantum circuit gradient estimation is a significant problem if VQF is to become competitive with classical factoring algorithms. Further, we compare entangled and non-entangled circuits in VQE optimization and fail to see significant evidence in favour of including entanglement in the VQE circuit.
\end{abstract}


\maketitle

\section{Introduction}

Factoring large composite integers is a textbook example of a computational problem that is hard to solve on a classical computer, the best known algorithms run in super-polynomial time. It is an immensely important problem since the RSA cryptosystem and therefore much of the information security of the modern world relies on its hardness.

One of the most famous applications of quantum computing is Shor's algorithm for integer factoring and computing discrete logarithms. It is one of the first known examples of superpolynomial advantage of quantum computers over classical computers \cite{Shor:1994}. When a sufficiently powerful quantum computer is built, that computer can be used to crack the security of most public key cryptosystems in use today. However, such a computer will not exist in the near future because it is estimated that tens of millions of qubits would be needed to factor cryptographically relevant integers with Shor's algorithm \cite{Gidney:2021}. It is estimated that around ten thousand qubits would be enough to factor 2048-bit RSA integers if quantum memory was available for the computation \cite{Gouzien:2021}. The large qubit requirement is due to the noise in current quantum computers. This noise can be dealt with by using quantum error correcting codes but unfortunately this requires too many qubits to be practically used in the near future.

Currently, the largest number factored with Shor's algorithm is 21 and even there the computation was simplified with prior knowledge of the Solution \cite{Lucero:2012,Smolin:2013}. Here, we study the performance of a heuristic alternative to Shor's algorithm suitable for Noisy Intermediate Scale Quantum (NISQ) computers, the so-called Variational Quantum Factoring (VQF) algorithm \cite{Anschutz:2017}. This suggested approach maps the factoring problem into the problem of finding a ground state of an Ising Hamiltonian and uses a quantum algorithm for finding this ground state.

Variational quantum algorithms (VQAs) are hybrid algorithms where a classical algorithm uses the quantum processor as a specialized hardware device to execute quantum circuits designed by the classical algorithm. This variational approach has been popular in recent years due to its ability to use noisy qubits more effectively: often only short quantum circuits with small numbers of qubits are needed and the effect of noise is therefore more manageable. They are also capable of mitigating some effects of noise and are flexible enough to match native quantum processor architectures and gate sets. There have been many proposed applications for VQAs, such as finding ground/excited states, simulation of quantum dynamics, combinatorial optimization, solving systems of equations, and machine learning algorithms \cite{Cerezo:2020}. The downside of VQAs is that in general, the classical optimization of circuit parameters is NP-hard \cite{Bittel:2021}.

We study the performance of different optimization algorithms in VQF and make resource estimates to see if we can find evidence of favourable scaling properties with increasing problem size. Currently, only the Quantum Approximate Optimization Algorithm has been applied to this problem. We find performance advantages in using a Variational Quantum Eigensolver instead. Additionally, we study the role of entanglement in the quantum circuit ansatz used in VQF.

\section{Variational quantum factoring}

Variational quantum factoring (VQF) is a proposed variational approach to integer factoring \cite{Anschutz:2017}. It works by encoding the factoring problem to the ground state of an Ising Hamiltonian. This ground state is then found with a VQA. Consider factoring $m = p q$ where $p$ and $q$ are assumed to be prime. In a binary representation, $m_{n_m}\ldots m_1 m_0 = p_{n_p} \ldots p_1 p_0 \cdot q_{n_q} \ldots q_1 q_0$. This equation implies a system of equations over the unknown binary variables $\{p_i\}$ and $\{q_i\}$. The equations are
\begin{align}
	m_i =& \sum_{j=0}^i q_j p_{i-j} + \sum_{j=0}^i z_{j,i} - \sum_{j=1}^{n_c} 2^j z_{i,i+j} \\
  \rightarrow C_i :=& \sum_{j=0}^i q_j p_{i-j} + \sum_{j=0}^i z_{j,i} - m_i - \sum_{j=1}^{n_c} 2^j z_{i,i+j} \nonumber \\
  =& \ 0 \label{eq:clauses}
\end{align}
where $i = 0, \ldots, n_c$ and $z_{i,j}$ are the carry bits originating from the binary multiplication. These equations are quantized to a Hamiltonian by the replacement
\begin{align}
	b_k \mapsto \frac{1}{2} (1 - Z_{b,k}) \ ,
\end{align}
where $Z$ is the Pauli-Z operator, $b=\{p,q,z\}$, and $k$ is the bit index. This replacement quantizes the clauses $C_i \mapsto \hat C_i$ and produces a Hamiltonian $\hat H = \sum_i \hat C_i^2$ over qubits whose ground state (with zero energy) is in one-to-one correspondence with bit assignments which satisfy $m=p q$. Therefore, factoring is reduced to the problem of finding the ground state of $\hat H$.

\section{Preprocessing}

Typically however, an additional simplification step is performed on \eqref{eq:clauses}, where the ``obvious'' equations are immediately solved. For example, one could deduce that $x y = 1$ is equivalent to $x = y = 1$, which can eliminate some of the binary variables. This preprocessing can be carried out in different ways and to different degrees resulting in a final system of equations which depends on all of those details in addition to the plain equation $m = p q$. Some factoring instances get significantly simplified with this process, sometimes instances might even get completely solved. The constraint this preprocessing step needs to satisfy is that it must run in polynomial time. More details on preprocessing can be found in \cite{Anschutz:2017}.

\section{Variational quantum eigensolver}

The variational quantum eigensolver (VQE) is a generic method for approximating ground states of Hamiltonians with parametrized quantum circuits (PQCs) \cite{Peruzzo:2014,Tilly:2021}. Since the factoring instance has been converted to a ground state problem, we can use VQE to solve factoring. The primary decisions to be made are the choice of the PQC that produces $\ket{\psi(\theta)}$ and the classical optimization method for $\theta$. For the classical optimization, we use BFGS algorithm which is a gradient-based optimization method where we compute the gradients with the parameter shift rule \cite{Guerreschi:2017,Mitarai:2018}.

We will compare two variational circuits in VQE. First we use a circuit with layers of $R_y(\theta)$-rotations separated by layers of controlled-$X$ operations. The size of the ansatz is controlled by the number of qubits $n$ and the number of $CX$-layers $L$.

\begin{figure*}
  \begin{quantikz}
    \lstick{$\ket 0$} & \gate{R_y} & \ctrl{1} & \qw & \targ{} & \gate{R_y} & \qw \rstick[wires=3]{$\ket{\psi(\theta)}$} \\
    \lstick{$\ket 0$} & \gate{R_y} & \targ{} & \ctrl{1} & \qw & \gate{R_y} & \qw \\
    \lstick{$\ket 0$} & \gate{R_y} & \qw & \targ{} & \ctrl{-2} & \gate{R_y} & \qw
  \end{quantikz}
  \hspace{2cm}
  \begin{quantikz}
    \lstick{$\ket 0$} & \gate{R_y} & \gate{T} & \gate{R_y} & \qw \rstick[wires=3]{$\ket{\psi(\theta)}$} \\
    \lstick{$\ket 0$} & \gate{R_y} & \gate{T} & \gate{R_y} & \qw \\
    \lstick{$\ket 0$} & \gate{R_y} & \gate{T} & \gate{R_y} & \qw
  \end{quantikz}
  \caption{The two variational circuits we use in our experiments. \textbf{Left:} One layer of the $CX$-circuit. Each $R_y$-gate is parametrized by an independent real number in the range $[0, 2\pi)$ representing the rotation angle. \textbf{Right:} One layer of the $T$-circuit. The difference to the $CX$-circuit is that the entangling $CX$-gates are removed in favour of single qubit $T$-gates. These circuits for $L=0$ mean including only the initial $R_y$-gates, in which case the $CX$ and $T$-circuits are equivalent.}
  \label{fig:circuits}
\end{figure*}
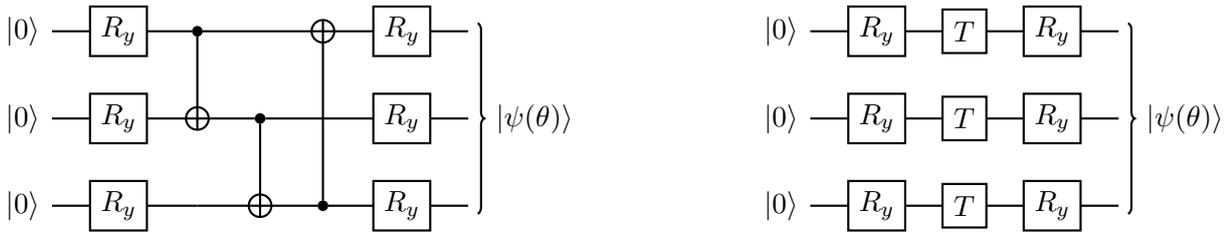

The second type of circuit we study is the first type of circuit with controlled-$X$ operations replaced by $T$ gates. This circuit clearly produces only product states and therefore can be efficiently simulated classically. The motivation for this is to study the role of entanglement in the experimental performance of VQE in integer factoring. It has been observed that at least in some cases, non-entangled circuits can perform similarly in some variational optimization tasks \cite{Nannicini:2019}. Without adding $T$-gates in between $R_y$-gates the rotations could simply be combined into one $R_y$-gate which would introduce redundancies in the set of variational parameters. Now since $[R_y(\theta), T] \neq 0$ when $\theta$ is not a multiple of $2\pi$, the different variational parameters acting on the same quantum wire still have non-trivial effects. If no performance difference is observed between $CX$- and $T$-circuits, then the experiment shows no signs of quantum advantage. The circuits have a repeating layered structure. The performance of VQF is controlled by changing the number of these layers. More layers means a more flexible state $\ket{\psi(\theta)}$ so finding the ground state is easier, but it also takes more time because there are more parameters to optimize. These circuits are shown in Fig.~\ref{fig:circuits}.

\section{Quantum approximate optimization algorithm}

The quantum approximate optimization algorithm (QAOA) is an algorithm inspired by a Trotterized version of quantum adiabatic optimization intended to solve combinatorial optimization problems \cite{Farhi:2014}. The circuit is a so-called quantum alternating operator ansatz
\begin{align}
  \ket{\psi(\gamma,\beta)} = \prod_j^p e^{-i \beta_j H_M} e^{-i \gamma_j H} \ket{+}^{\otimes n} \ ,
\end{align}
where $H_M = \sum_i^n X_i$ is the mixing Hamiltonian. The initial state $\ket{+}^{\otimes n}$ is an eigenstate of $H_M$, which is supposed to be transformed to a ground state of the problem Hamiltonian $H$ by a good choice of parameters $\beta$ and $\gamma$, analogously to quantum adiabatic optimization. The quantum processor is used for computing $E = \expval{H}{\psi(\gamma,\beta)}$ and the parameters $\beta$ and $\gamma$ are optimized classically. The algorithm gives increasingly better results for increasing the number of parameters $p$. QAOA unitaries are shown to be universal for quantum computing \cite{Morales:2020}. The downside is that the circuit length is linear in the number of terms in $H$ resulting in circuits that quickly become too long for reliable evaluation on NISQ processors.

\section{Resource requirements}

In the theory of computational complexity and analysis of algorithms, two fundamental resources are of concern: time and space. The time required by an algorithm is measured in the number of elementary operations it needs to produce a result. Space requirement is measured as the minimum number of bits required to execute the algorithm. In order to measure the performance of VQF, we need to decide what we consider an elementary quantum operation in the algorithm. In this work, we will consider both the number of quantum circuit shots and quantum gates. These numbers are related as can be seen by noticing that each of our circuits contain $\mathcal O (L n)$ gates. We'll also ignore all other time consuming parts in the quantum algorithm, including all related to classical processing. All these simplifications tip the scales in the quantum algorithm's favour. We are also ignoring that quantum gates are orders of magnitude slower than classical operations \cite{Babbush:2020}. This motivates the choice to ignore classical operations, the runtime of realizing this algorithm is expected to be dominated by the quantum part.

For the optimization part of the algorithm, we will use BFGS which is a gradient-based optimizer. Unfortunately, such optimization methods do not have runtime guarantees for finding the minimum. In fact, there are no guarantees to find the global minimum at all. Therefore, we will start by analysing the time requirements of only one gradient evaluation because this is deterministic and the number of gradient evaluations needed to converge is not known ahead of time. Later we will run experiments to estimate how many steps one needs to converge in BFGS. Now, we will relate the parameters $n$ and $L$ to the number of shots $N_{shots}$ and the number of gates $N_{gates}$ required in one gradient evaluation.

The number of qubits is an important bottleneck for NISQ devices. Factoring larger numbers often requires more qubits in the VQF algorithm so the number of qubits places a practical bound on the size of numbers to be factored. If preprocessing of the equation system is used, the relationship between input number size $n_m$ and the number of qubits required $n$ is complicated, but the rule still holds that often a larger number requires more qubits in VQF. In addition, in \cite{Anschutz:2017} prior knowledge of $n_p$ and $n_q$ was assumed, which obviously is not possible in real factoring challenges. This prior knowledge substantially reduces qubit requirements.

\begin{figure}
  \centering
  \includegraphics[width=0.45\textwidth]{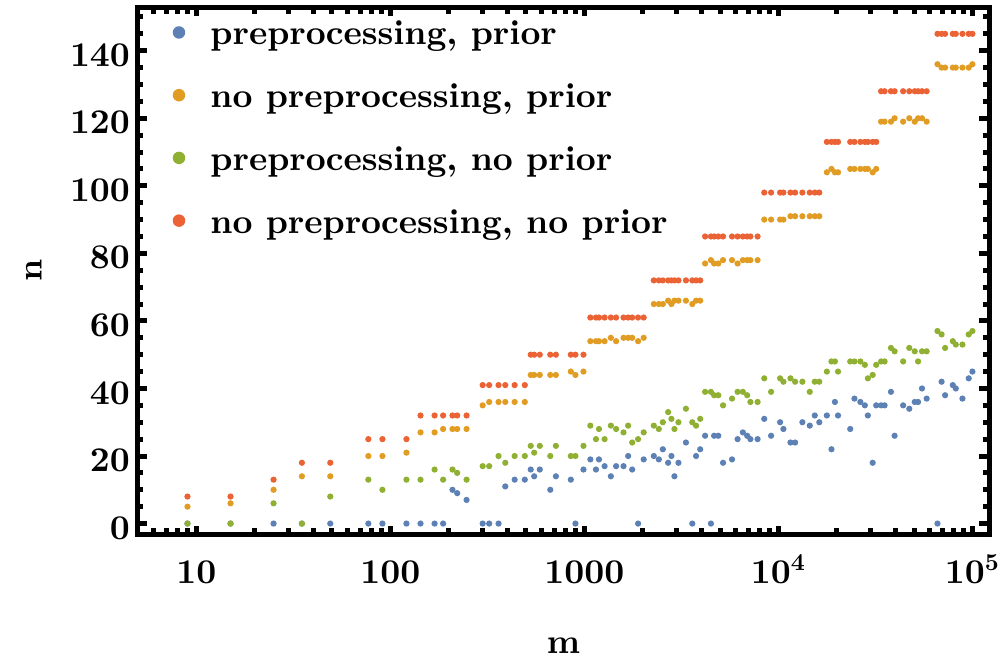}
  \caption{The number of required qubits to factor biprimes of different sizes. Different points correspond to qubit numbers with/without preprocessing and with/without prior knowledge of $n_p$ and $n_q$.}
  \label{fig:qubit_numbers}
\end{figure}

In Fig.~\ref{fig:qubit_numbers} we show the dependence of the required number of qubits $n$ on whether or not preprocessing or prior knowledge of $n_p$ and $n_q$ is used. It can be seen that these choices have a significant effect on $n$. The realistic case is where one uses preprocessing without unfair prior knowledge. In both preprocessed cases, we observe approximately linear dependence of the qubit number on $n_m = \lceil \log m \rceil$. If we extrapolate under the linear assumption, we find that in the case with no prior knowledge we would need around 9300 qubits to represent a factoring instance of a 2048-bit integer.

\begin{figure}
  \centering
  \includegraphics[width=0.45\textwidth]{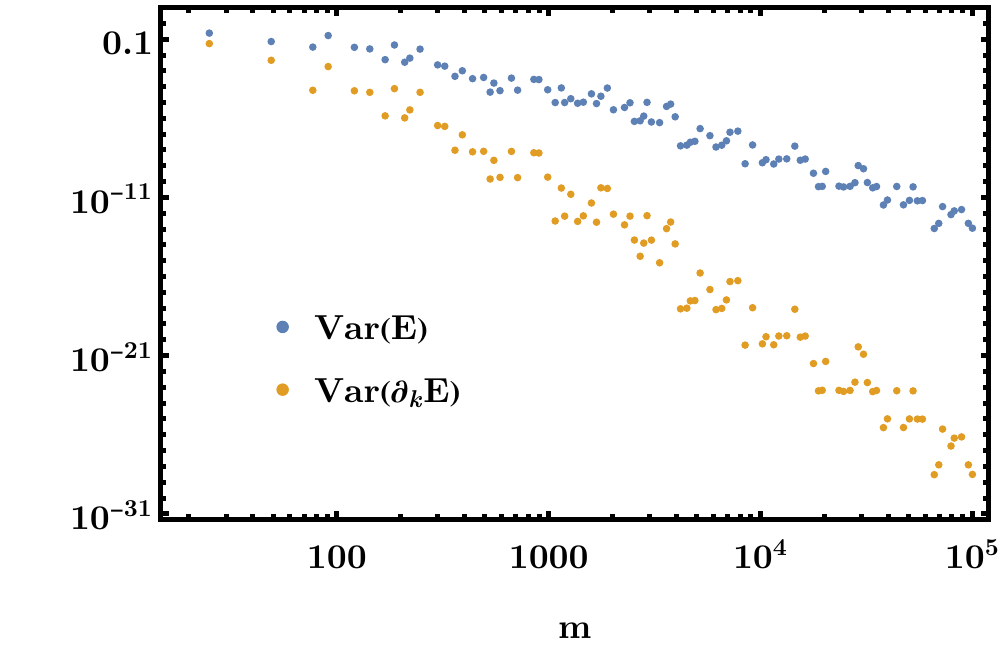}
  \caption{The variance of $H$ for factoring instances of various integers $m$. All instances are preprocessed without prior information.}
  \label{fig:variance}
\end{figure}

Since the Hamiltonian is diagonal, we can measure all Pauli terms simultaneously. In order to estimate one expectation value to precision $\epsilon$, we need to run
\begin{align}
  N_{shots} = Var(\expval{H}{\psi(\theta)}) / \epsilon^2 \ ,
\end{align}
shots on the quantum processor. The expectation value $E(\theta) = \expval{H}{\psi(\theta)}$ depends on the parameters $\theta$, but for simplicity we'll drop the $\theta$-dependence by averaging over the states $\ket\psi$ w.r.t. the Haar measure on the Hilbert space. Thus,
\begin{align}
  & \mathbf E (\expval{H}{\psi(\theta)}) \approx \int \bra 0 U^\dagger H U \ket 0 \dd U \\
  & = \frac{\tr H}{2^n} \quad \text{and} \\
  & \text{Var}(\expval{H}{\psi(\theta)}) \\
  & \approx \int \qty( \bra 0 U^\dagger H U \ket 0 )^2 \dd U - \mathbf E (\expval{H}{\psi(\theta)})^2 \\
  & = \frac{2^n \tr H^2 - \qty( \tr H )^2}{2^{2n}\qty(2^n + 1)} \ .
\end{align}
These formulas are for estimating the values of the energy itself. However, we are also interested in the gradient of energy because this is needed for optimization. For the $k$th component of the gradient of $E = \expval{H}{\psi(\theta)}$ we have
\begin{align}
  & \mathbf E (\partial_k E) = 0 \label{eq:grad_exp}\\
  & \text{Var}(\partial_k E) = \frac{\tr H^2}{2^{3n - 2}} \label{eq:grad_var} \ ,
\end{align}
assuming that the parametrized gates are generated by Pauli matrices, which is the case in our circuits \cite{McClean:2018}. Under these assumptions the derivatives w.r.t. any parameter in the quantum circuit have the same statistical properties as can be seen by noting the absence of $k$ on the right hand side of \eqref{eq:grad_exp} and \eqref{eq:grad_var}. Fig.~\ref{fig:variance} shows this variance experimentally as a function of the input integer $m$. Variances are computed with preprocessing but without prior knowledge about $n_p$ and $n_q$. Given that Fig.~\ref{fig:qubit_numbers} shows approximately linear dependence between $n$ and $\log m = n_m$, implies that the variances in Fig.~\ref{fig:variance} decay exponentially in $n_m$.

We will use gradient based optimization in VQE. Both $CX$ and $T$-circuits have $(L + 1) n$ parameters. Then for each energy gradient evaluation, we need to estimate $(L + 1) n$ individual derivatives of an expectation value. This means, that we need to run
\begin{align}
  N_{shots} = (L + 1) n \frac{\text{Var}(\partial_k E)}{\epsilon^2} \label{eq:N_shots}
\end{align}
shots to estimate one gradient to precision $\epsilon$. Counting the number of quantum gates we need for each gradient, we obtain
\begin{align}
  N_{gates} = (L + 1)(2 L + 1) n^2 \frac{\text{Var}(\partial_k E)}{\epsilon^2} \ , \label{eq:N_gates}
\end{align}
since the VQE circuit contains $(L + 1) n$ single qubit rotations and $L n$ controlled NOT-gates.

We want to compare the performance with classical factoring algorithms. We choose to test against the simplest possible algorithm, trial division. The algorithm starts at $p=2$, checks if $p$ divides $m$, and increments $p$ until a divisor is found. At the worst case, this requires $\lfloor \sqrt m \rfloor - 1$ divisions since we must have $p \leq \sqrt m$. We will not consider space requirements since the classical bottleneck is the available time, not the memory.

\begin{figure}
  \centering
  \includegraphics[width=0.56\textwidth]{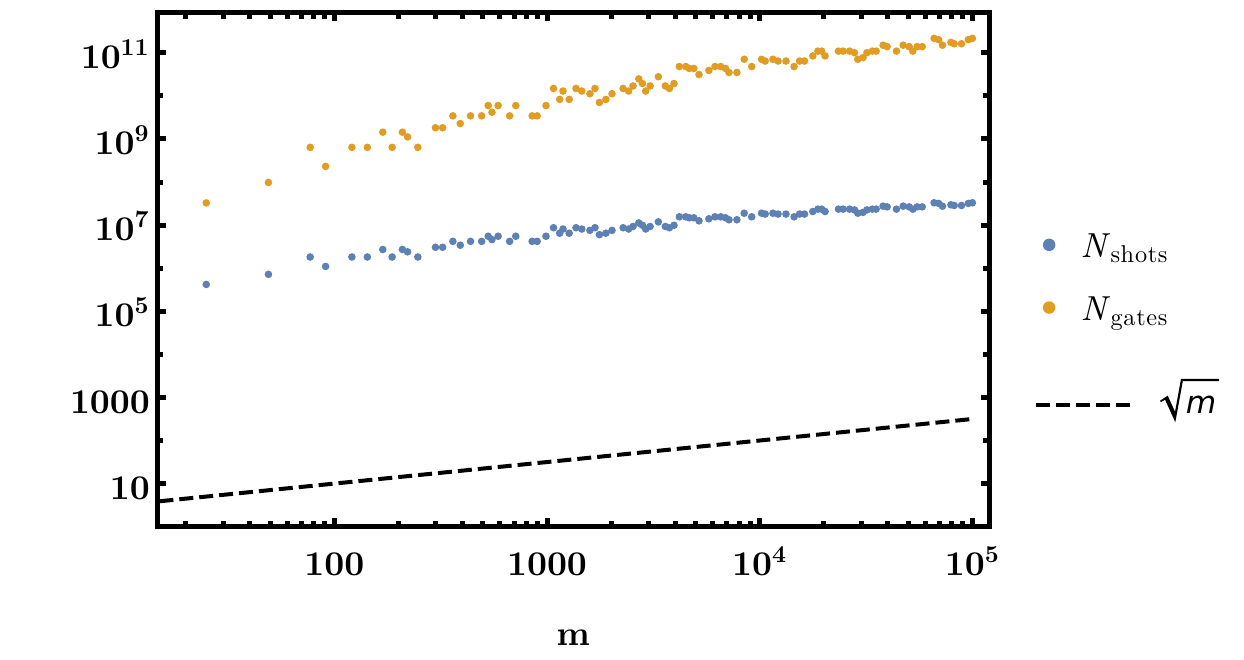}
  \caption{The number of shots $N_{shots}$ and the number of gates $N_{gates}$ one needs to execute to estimate one gradient of $E(\theta)$, from eqs. \eqref{eq:N_shots} and \eqref{eq:N_gates} for different factoring instances $m$. We consider $CX$-circuits with $L = n$. The precision parameter is set to $\epsilon = \sqrt{\text{Var}(\partial_k E)}/100$, this way most circuits are estimated to precision better than one standard deviation. The dashed line denotes $\sqrt m$, the maximum number of trial divisions needed to factor $m$.}
  \label{fig:shots_and_gates}
\end{figure}

It is interesting to study how $N_{shots}$ and $N_{gates}$ behave for different factoring instances $m$. We plot these quantities for various $m$ in Fig.~\ref{fig:shots_and_gates}. We see that for factoring instances $m$, trial division uses orders of magnitude less divisions comparing to the number of circuit shots or quantum gates needed to estimate a single circuit gradient. Note that the dashed curve representing trial division is the number of divisions needed to factor $m$ deterministically, in the worst case. The VQF algorithm on the other hand will need to evaluate many gradients in order to optimize parameters so that the ground state, and equivalently, $p$ and $q$ are found.

\section{Solution manifold}

Typically in VQF one is interested in finding the ground state of the problem Hamiltonian. The ground state after all corresponds to the factors $p$ and $q$ by construction. This is what our optimization is trying to do as well, but one could object that finding the exact ground state is an unnecessarily strict requirement in practice. What actually suffices is to find either $p$ or $q$ during optimization while ignoring the values of the carry bits $z$. Optimization algorithms could be seen as a fancy version of trial division where candidate divisors are sampled from a probability distribution defined by the ansatz state $\ket{\psi(\theta)}$. One could imagine periodically estimating the energy $E(\theta)$ during optimization. That estimate would require sampling $\ket{\psi(\theta)}$ and those samples could be checked classically. The ``solution manifold'' was defined in \cite{Anschutz:2017} to describe the part of the state space which corresponds to correct assignment for $p_i$ and $q_i$ but excludes the $z_{i,j}$-bits. We will consider a more relaxed version of this concept: in this paper, the solution manifold is the set of computational basis states which corresponds to the correct assignments of either $p_i$ or $q_i$-bits. If energy is sampled during optimization, we could finish at the moment when any bit string in the solution manifold is sampled and classically checked to divide $m$. Therefore the size of this solution manifold is clearly of interest.

\begin{figure}
  \centering
  \includegraphics[width=0.45\textwidth]{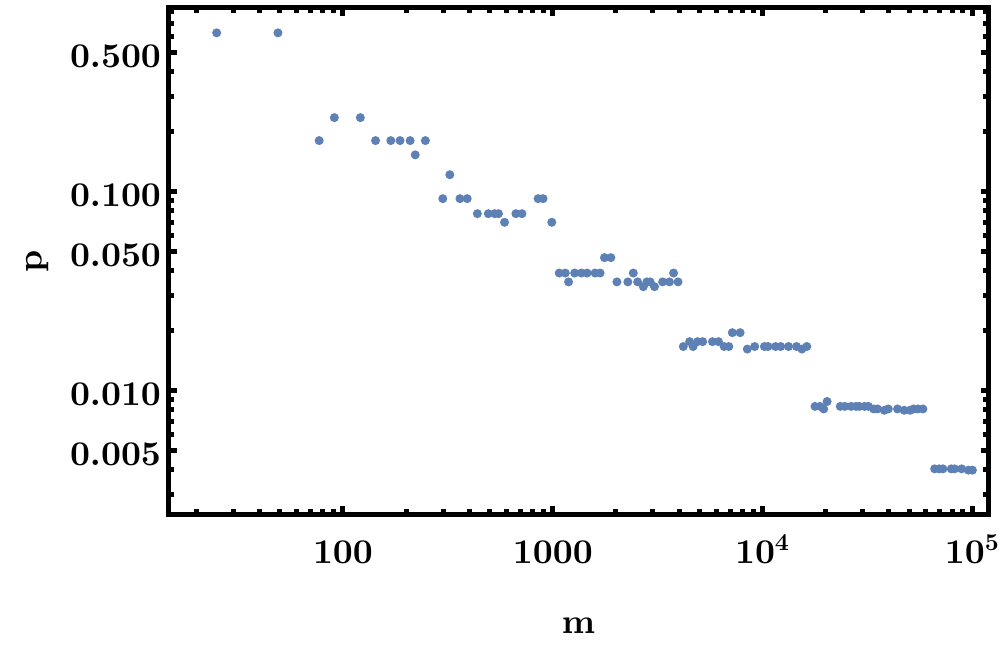}
  \caption{The fraction of all computational basis states that belong to the solution manifold, for various $m$. Each instance is preprocessed without prior knowledge. The solution manifold size decreases exponentially in $\log m$.}
  \label{fig:sol_manifold}
\end{figure}

Fig.~\ref{fig:sol_manifold} shows the relative size of the solution manifold to the number of computational basis states. One can see that the solution manifold contains a large fraction of the computational basis states for small $m$ but this fraction diminishes quickly as $m$ increases. The problem with the usefulness of the solution manifold is that the algorithm is optimizing the energy of the ansatz state, not the overlap with the solution manifold. The result is that often the optimizer reduces the probability of observing a solution manifold state when sampling $\ket{\psi(\theta)}$ at least until the ground state is found, if it is found. This can be understood by remembering that the optimization process is trying to increase the amplitude of low energy states, but these low energy states often do not lie in the solution manifold. For example, a state with only two incorrect bits is not in the solution manifold if one of those bits appears in $p_i$ and the other in $q_i$. The average energy of solution manifold states is also only marginally lower than states outside of the manifold. For example, in the case of $m=91$ the average solution manifold state energy is $15.9$. The average energy of states outside the manifold is $17.8$, which is not that much larger. Therefore we do not expect the solution manifold to help much with practical performance. If the optimizer happens to find the ground state, then the probability of observing a solution manifold state will approach 1. However, if the optimizer gets stuck in a local suboptimal minimum, then the probability likely tends to zero.

\section{Results}

\subsection{Optimization algorithm}

The algorithm for finding the ground state of $H$ is of vital importance for the performance of VQF. The popular choice in the literature for VQF and more generally for combinatorial optimization problems is QAOA. Therefore we chose to compare the performance between QAOA and VQE. One immediate advantage of VQE is that it typically uses considerably shorter circuits than QAOA. The QAOA ansatz circuit length strongly depends on $H$, and in our experiments the Hamiltonian can have hundreds of Pauli terms. This makes the QAOA circuits too deep to practically evaluate on the noisy quantum devices of the near future. The advantage of QAOA is that it is theoretically understood to find the ground state when the number of parameters becomes large.

\begin{figure}
  \centering
  \includegraphics[width=0.45\textwidth]{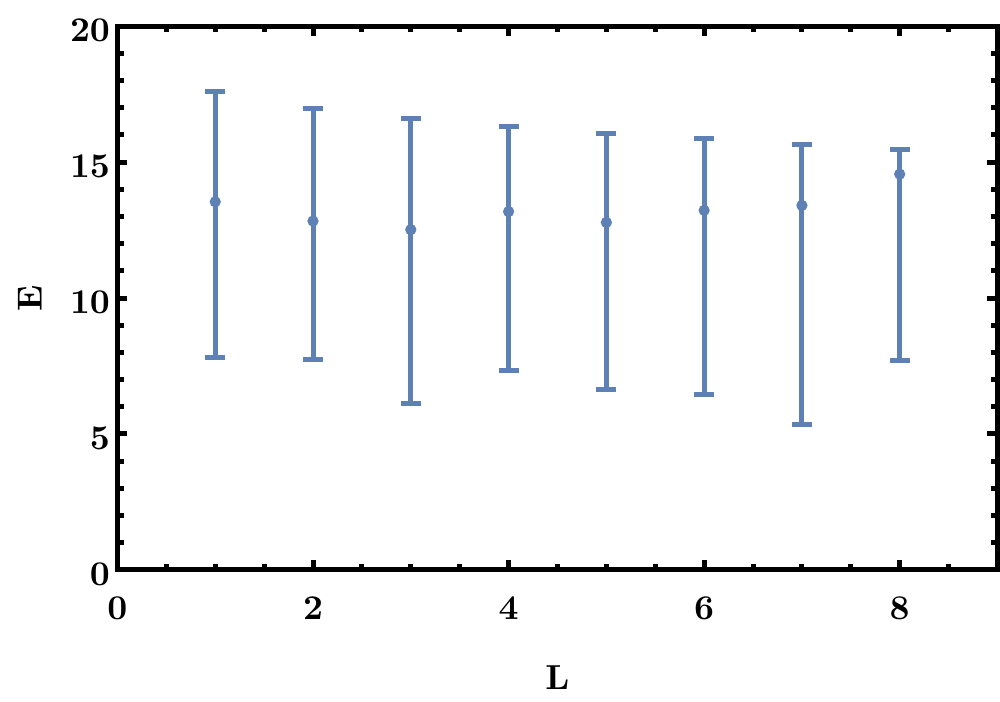}
  \caption{Final energies for $L$ layers of the QAOA ansatz optimized with BFGS. Average energies and error bars ($5\%$ and $95\%$ quantiles) are computed over $100$ random initializations. The factoring instances for different $m$ use classical preprocessing without prior knowledge.}
  \label{fig:final_energy_m_91_qaoa}
\end{figure}

We find that VQE performs considerably better than QAOA in factoring problems. The final energies of QAOA and VQE are shown in Figs.~\ref{fig:final_energy_m_91_qaoa} and \ref{fig:final_energy_m_91}, respectively. Both algorithms are run 100 times with random initializations of their parameters for the classical BFGS optimizer. The factoring instance in question is $m=91$ which corresponds to 10 qubits with preprocessing but without prior knowledge of $n_p$ or $n_q$. It is clear that VQE reaches lower energies than QAOA and does it consistently with shorter circuits. The trade-off is that the VQE circuits have considerably more free parameters to optimize than QAOA.

For either algorithm, the number of layers does not seem to have a radical effect on the final energies. Both QAOA and VQE become more expressive when the number of parameters increases. The error bar gets slightly lower for QAOA when $L$ increases. For VQE, there is no significant difference in performance after the first couple of layers are added.

\subsection{Entanglement}

\begin{figure}
  \centering
  \includegraphics[width=0.45\textwidth]{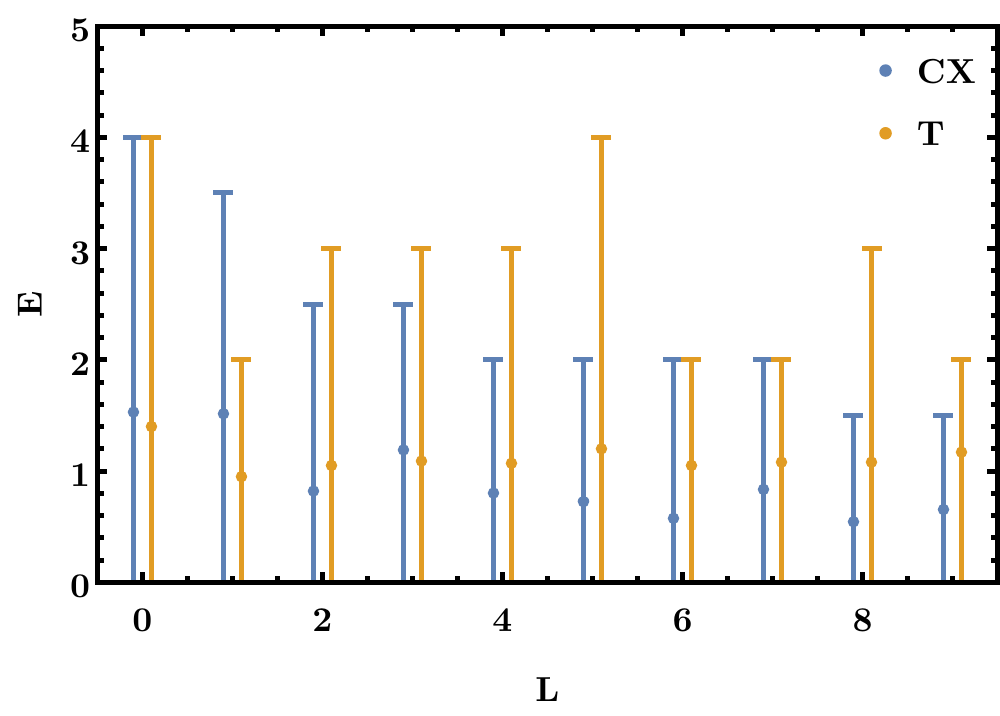}
  \caption{Average final energy for $m=91$ ($10$ qubits) when using VQE optimized with BFGS. Average energies and error bars ($5\%$ and $95\%$ quantiles) are computed over $100$ random initializations. This factoring instance uses preprocessing but does not use prior knowledge. We see that there is no significant difference in performance between $CX$ and $T$-circuits.}
  \label{fig:final_energy_m_91}
\end{figure}

Since we have reformulated the factoring problem as a combinatorial optimization problem, it is not clear that entanglement is required in the ansatz state. That is, the ground state is by construction a computational basis state which is unentangled and easy to produce with a parametrized circuit without connections between qubits. This follows from the classicality of the computational problem, entanglement would obviously be a necessity in any optimization problem where the ground state is expected to be entangled. However, it seems possible that introducing entanglement might speed up convergence to the ground state or increase the probability of finding the ground state. An entangled circuit might offer these speedups by being able to move from the initial state to the ground state via entangled states of the Hilbert space. An unentangled circuit would have to move in the state space along more constrained trajectories, only passing through product states. Also if there is no performance improvement in using entanglement, then there is also no quantum advantage: product states on $n$ qubits can be represented and computed with polynomial amount of classical resources. It is experimentally known that at least in some cases entangled circuits do not seem to offer performance advantages \cite{Nannicini:2019}. This is also the case in our experiments. Fig.~\ref{fig:final_energy_m_91} shows the final optimization energy of 100 random BFGS initializations for both entangled $CX$ and unentangled $T$-circuits. The entangled circuits seem to have a slightly better chance of finding lower energy states as can be seen from the lower error bars in Fig.~\ref{fig:final_energy_m_91} but this effect seems quite weak. Also the entangled circuits have lower average final energies but the difference to the unentangled circuit is small. It would be interesting to see similar experiments in larger instances or other optimization problems since at least here we do not see any strong evidence of quantum advantage in this kind of optimization problem. Also one could ask whether a different choice of quantum circuit would change these results. For example, one could ask whether adding $R_x$-rotations would make a difference.

\subsection{Instance size}

\begin{figure}
  \centering
  \includegraphics[width=0.45\textwidth]{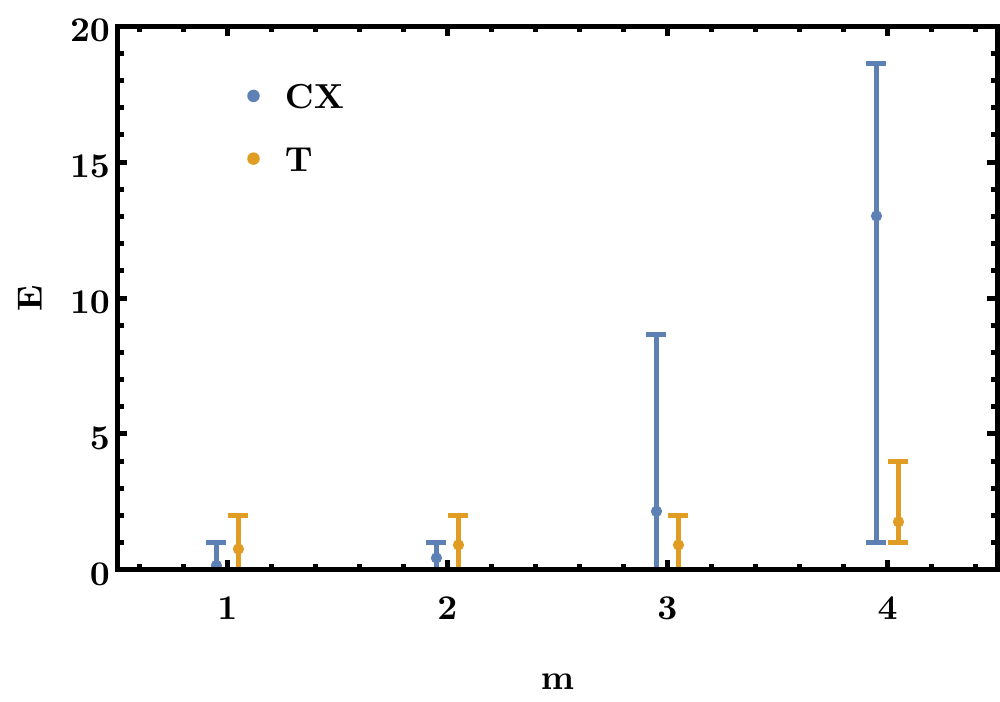}
  \caption{Final optimization energies for $m = 25$, $49$, $91$, and $247$. Average energies and $5\%/95\%$ quantiles are once again computed over 100 random initializations in BFGS. We are using both $CX$ and $T$-circuits with $N=n$ layers for each instance. These instances use classical preprocessing but do not use any prior knowledge.}
  \label{fig:vqe_final_energy_vs_m}
\end{figure}

The scaling of performance with respect to the problem size is an important piece of information when comparing algorithms. Since we cannot analytically study the asymptotic scaling properties of VQF, we must resort to numerical experiments. This is however also somewhat unsatisfying because we are restricted to very small factoring instances when simulating the algorithm on a classical computer. Our simulation results are shown in Fig.~\ref{fig:vqe_final_energy_vs_m}. We are restricted to studying the biprimes $25$, $49$, $91$, and $247$. These correspond to $6$, $8$, $10$, and $13$ qubits, respectively. In all instances, for optimization we use VQE with $L=n$ layers for each type of ansatz circuit. It is interesting to see that the performance of the non-entangled $T$-circuit doesn't suffer nearly as much as the $CX$-circuit's performance as $m$ increases. Both circuits can achieve the same minimum energies but there is a lot more variance with the entangled circuit signalling a higher difficulty in traversing through the energy landscape successfully.

\subsection{Steps until convergence}

\begin{figure}
  \centering
  \includegraphics[width=0.45\textwidth]{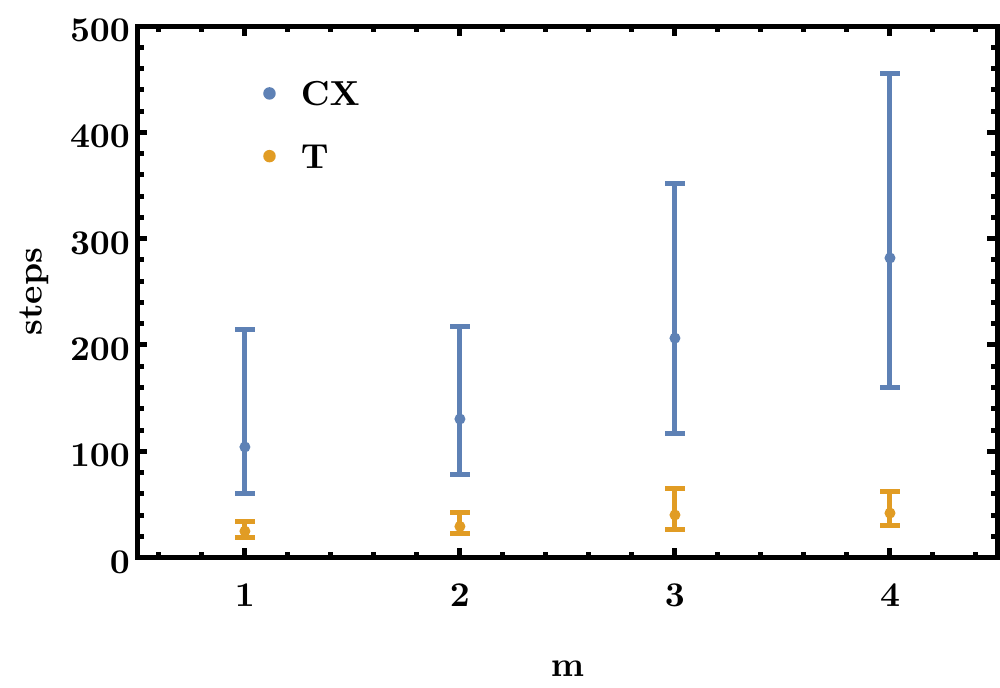}
  \caption{The number of gradient steps needed for convergence in BFGS optimization of VQE for different $m$. The VQE circuit has as many layers as qubits in each instance. That is, there are $6$, $8$, $10$, and $13$ qubits/layers, respectively.}
  \label{fig:vqe_steps_vs_m}
\end{figure}

\begin{figure}
  \centering
  \includegraphics[width=0.45\textwidth]{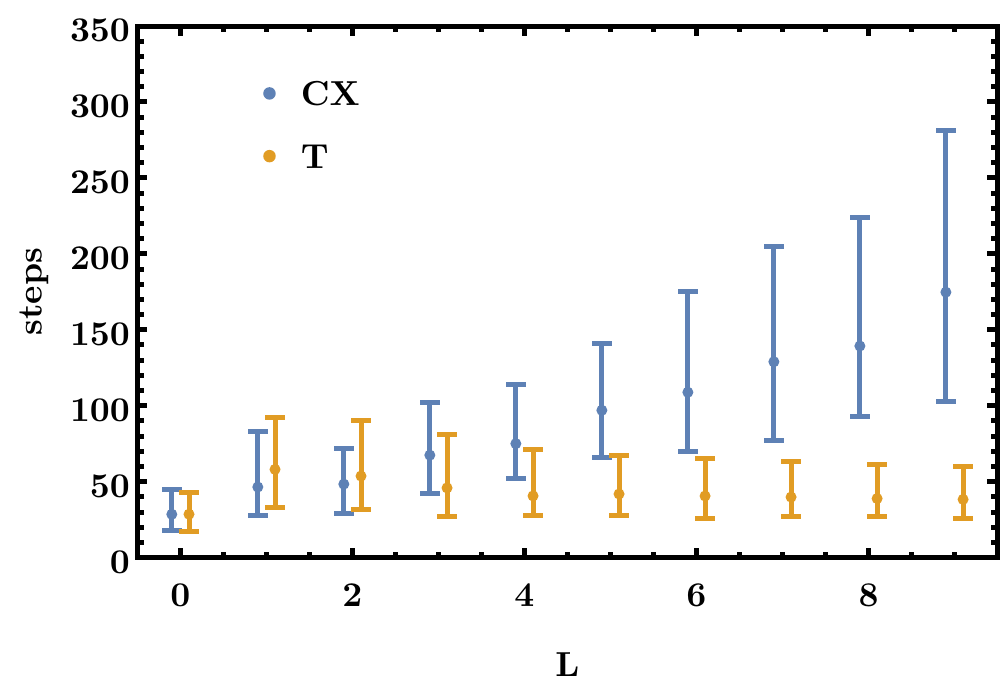}
  \caption{The number of gradient steps until BFGS convergence, this time we fix $m=91$ and vary both the number of layers $L$ and the type of the VQE circuit.}
  \label{fig:vqe_m_91_steps_vs_L}
\end{figure}

Previously we made estimates for the number of quantum circuit shots and quantum gate executions needed to compute one gradient in the parametrized quantum circuit of VQE. When studying an optimization algorithm's performance, an equally important quantity is the number of steps needed to converge. Again, this is not something we can derive analytically so we must resort to numerical simulations. In Fig.~\ref{fig:vqe_steps_vs_m} we show the typical number of gradients needed to evaluate before BFGS converges. We can see that the number of gradients increases as a function of $m$ for the $CX$-circuit. The required number of gradients increases also for $T$-circuits but much more slowly. It seems that in this optimization problem, $T$-circuits perform at least as well as $CX$-circuits and are also faster in two ways: they need less gradient steps to achieve convergence and are efficient to simulate classically.
The number of gradient steps as a function of the number of layers $L$ in the ansatz is shown in Fig.~\ref{fig:vqe_m_91_steps_vs_L} for the case of $m=91$ for both $CX$ and $T$-circuits. It can be seen that the needed number of gradients to evaluate increases when the instance size increases, as well as for increasing number of layers in the case of the $CX$-circuit. The $T$-circuit on the other hand does not see a similar increase in the number of gradient steps when $L$ increases, even though the average energy reached is not significantly different from those achieved by the $CX$-circuit.

\section{Conclusions}

In this paper we studied the performance of the variational quantum factoring algorithm. Since the algorithm is heuristic, performance has to be characterized experimentally. The algorithm has two major moving parts: the process which encodes the factoring problem to a ground state of a Hamiltonian and the quantum optimization algorithm used for extracting that ground state. We found that the number of qubits the Hamiltonian needs to act on depends quite strongly on whether or not classical preprocessing is used on the initial clauses. Also the effect of prior information is significant. This prior information about the factors of $m$ has been used in some previous papers in order to reduce the qubit requirements of the algorithm so simulations are easier to run on classical hardware. We also produced estimates for the number of shots/gates needed to evaluate one gradient in any gradient-based optimization algorithm and found that gradient evaluation is quite resource intensive. Therefore the time required for quantum circuit gradient evaluation overwhelms the time required for complete classical factoring for the range of $m$ we studied. We compared two quantum algorithms for ground state finding in VQF and found that VQE consistently achieves lower energies than QAOA for our factoring instances. Further, we studied two different parametrized circuits in VQE, one with entanglement and one without. We found that the entangled circuit found similar final energies as the unentangled circuit on average for $m=91$. On the other hand, the unentangled circuits seem to have better average performance as $m$ increases. However, the unentangled circuits converged substantially faster than the entangled circuits when their parameters were optimized with BFGS. We take this as an indication for the need to go beyond problem-agnostic hardware-efficient ansatz circuits in quantum optimization and taking advantage of the structure of the specific problem under consideration.

\section{Acknowledgements}
This research was supported by the Post-Quantum Cryptography project funded by Business Finland's Digital Trust program, the Flagship on Photonics Research and Innovation (PREIN) programme managed by
the Academy of Finland, and the Finnish Quantum Institute (InstituteQ).

\bibliographystyle{quantum}
\bibliography{biblio}

\end{document}